\documentclass[draftclsnofoot,onecolumn,12pt]{IEEEtran}
\usepackage{multirow}
\usepackage{booktabs}
\usepackage{amsfonts}
\usepackage{amsmath,cases}
\usepackage{amssymb}
\usepackage{graphicx,graphics}
\usepackage{stfloats}
\usepackage{subfig}
\usepackage{epstopdf}
\usepackage{ifpdf}
\usepackage{cite}
\usepackage{epsfig}
\usepackage{verbatim}
\usepackage{balance}
\usepackage{url}
\usepackage{algorithm}
\usepackage{algorithmicx, algpseudocode}
\usepackage{color}

\newcommand{\bp}{\mathbf{p}}

\begin{document}
\title{Real-time Optimal Resource Allocation for Embedded UAV Communication Systems}
\author{Minh-Nghia Nguyen, Long D. Nguyen, Trung Q. Duong, \\ and Hoang Duong Tuan
	\thanks{This work was supported by the Newton Prize 2017 and by a Research Environment Links grant, ID 339568416, under the Newton Programme Vietnam partnership.}
	\thanks{M.-N. Nguyen is with the Institute of Research and Development, Duy Tan University, Da Nang 550000, Vietnam (e-mail: m.nghia.n@gmail.com). L. D. Nguyen and T. Q. Duong are with Queen's University, Belfast BT7 1NN, UK (e-mail:\{lnguyen04, trung.q.duong\}@qub.ac.uk). H. D. Tuan is with the University of Technology Sydney, Broadway, NSW 2007, Australia (e-mail: tuan.hoang@uts.edu.au).}
}
\maketitle
\begin{abstract}
We consider device-to-device (D2D) wireless information and power transfer systems using an unmanned aerial vehicle (UAV) as a relay-assisted node. As the energy capacity and flight time of UAVs is limited, a significant issue in deploying UAV is to manage energy consumption in real-time application, which is proportional to the UAV transmit power. To tackle this important issue, we develop a real-time resource allocation algorithm for maximizing the energy efficiency by jointly optimizing the energy-harvesting time and power control for the considered (D2D) communication embedded with UAV. We demonstrate the effectiveness of the proposed algorithms as running time for solving them can be conducted in milliseconds.
\end{abstract}

\begin{IEEEkeywords}
Energy harvesting, energy efficiency, unmanned aerial vehicles, device-to-device network, real-time embedded optimization.
\end{IEEEkeywords}

\section{INTRODUCTION} \label{sec:Intro}
Unmanned aerial vehicle (UAV)-based communication networks with their flexible configuration and mobility nature can be more efficient and inexpensive for deployment of future wireless network \cite{Gupta2016survey} and the Internet of Things (IoT) applications \cite{MTA16}. Moreover, it has been emphasized that UAV-based wireless systems are capable of enhancing wireless communications by virtue of the dominant presence of line-of-sight (LOS) connections \cite{Gupta2016survey}. Therefore, UAVs can totally provide novel schemes to enhance the network coverage for serving more wireless devices. A major issue in UAV-based applications is that UAV devices typically have limited energy storage for flying operations whereby the deployment and resource allocation such as spectrum or transmit power allocation should be considered for efficient utility \cite{Zhang2017CommLett, Baek2017WireCommLett}. However, there are only a few existing works that concentrate on the resource allocation aspect to improve the energy efficiency (EE) performance of UAV-based networks \cite{ZZ17}.
	
Although UAV has been widely recognized as a promising technology to improve wireless networks performance, its fundamental potential has not fully been exploited.
An interesting development in UAV-based networks is the application of wireless energy transfer (WPT). As a matter of fact, WPT in radio frequency has recently promised advance technology for providing energy to wireless devices over the air (see e.g. \cite{BHZ2015} and the references therein).
Very recently, WPT for UAV-enabled device-to-device (D2D) networks has been considered in \cite{Wang2017TGCN} where UAVs can operate as an energy supplier for multiple D2D pairs.
Nevertheless, this work only considers throughput maximization and does not focus on the aforementioned EE problem, which is crucial for providing efficient and lasting operation.
To tackle this issue, we address the EE problem in the scenario of UAV-based relay network supporting energy harvesting-enabled D2D communications.
In particular, we investigate the issues of not only power allocation but also energy harvesting time, which will be formulated as a joint optimization problem in energy harvesting-powered D2D communications. Nevertheless, joint optimization problems are often complicated, for which we propose low-complexity efficient resource allocation algorithms.

Another critical issue in UAV is the real-time control and operation due to its lifetime and dynamic environment. As such, we study the resource allocation problem for energy harvesting-powered D2D communications underlaying UAV networks using real-time optimization. With the rapid improvement of computational speed as well as the use of efficient algorithms and advanced coding approaches, the embedded convex approach is able to solve the optimal resource allocation problems in the level of microseconds or milliseconds time scales with strict time limits \cite{MattingleySPM2010}. A lot of resource allocation optimization problems have been considered in wireless communications. However, there is still a lack of investigation for real-time optimization problem. With the exceeding development of computing performance, the solving optimization problems in real world have become a necessary trend of wireless communication. To fulfil this gap, for the first time, this paper has considered real-time optimization for resource allocation for embedded UAV-based  communication systems.
\begin{figure}[H]
	\centering
	\centerline{\includegraphics[trim=0cm 0cm 1cm 1cm, width=0.6\textwidth]{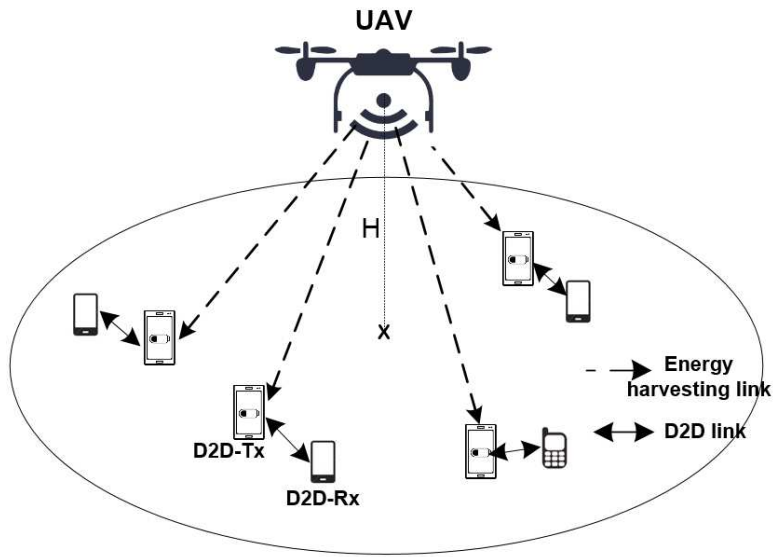}}
	\caption{D2D communications assisted by UAV.}
	\label{fig:system_model}
\end{figure}

\section{Problem statements}
Consider a communication system with one UAV mobile serving multiple energy harvesting-powered D2D pairs as shown in Fig. \ref{fig:system_model}. The UAV and users are equipped with a single antenna. Given a unitary communication time-slot, the energy harvesting and transmit information in UAV D2D network occur in two phases. In the first phase spanning $\tau$ with $0<\tau<1$, the dedicated D2D-transmitter (D2D-Tx) harvests energy from UAV. Then, in the second phase spanning $\left(1 - \tau\right)$ the information transmission happens between D2D pairs.
The set of D2D pairs is denoted by $\mathcal{N} = \{1, 2, ... ,N\}$. The energy harvested at the $n$th D2D-Tx is given by
\begin{equation} {\label{modl_1}}
E_n = \tau \eta P_0 g_n
\end{equation}
where $0 < \eta < 1$ is the energy harvesting efficiency, $P_0$ is the maximum total transmit power at UAV, and $g_n$ is the channel power gain from the UAV to $n$th D2D-Tx.

For practice purpose, each user is assumed to utilize the harvested energy for information transmission phase. Denote by $p_n$  the transmission power of the $n$th D2D pair. The following energy causality constraint must be satisfied
\begin{equation}\label{modl_2}
(1 - \tau) p_n \leq \tau \eta P_0 g_n, \ n \in \mathcal{N}
\end{equation}
For $\bp = [p_n]_{n=1}^N$, the information throughput (in nats) at $n$th D2D pair is
\begin{equation}\label{modl_3}
r_n(\tau, \bp) = (1 - \tau) \ln \left(1 + \frac{p_n h_{n,n}}{\sum_{i \neq n}^{N} p_i h_{n,i} + \sigma^2}\right)
\end{equation}
where $h_{n,i}$ is the channel gain for the link from the $n$th D2D-Tx  to $i$th D2D-receiver (D2D-Rx) and $\sigma^2$ is the noise power.
Next, the total power consumption in the considered D2D network is written by
\begin{equation}\label{modl_4}
\vartheta(\tau, \bp) = \sum_{n=1}^{N} (1 - \tau) p_n + \tau \eta P_0 + P_{\sf cir}.
\end{equation}
where $P_{\sf cir}$ is the circuit non-transmit power at the UAV.

In our work, the main target is to maximize the EE of UAV networks while satisfying the energy causality constraint and quality-of-service (QoS) constraint for each D2D pair. As such, the EE maximization problem is as follows:
\begin{subequations}\label{ee_modl}
\begin{align}
\max_{\tau, \bp > 0} & \ \phi = \frac{\sum_{n=1}^N r_n(\tau, \bp)}{\vartheta(\tau, \bp)} \label{ee_modl_a} \\
\mbox{s.t.} & \ (\ref{modl_2}), \\
& \ r_n(\tau, \bp)\geq \bar{r}, n \in \mathcal{N},\label{ee_modl_c} \\
& \ 0 \leq \tau \leq 1, \label{ee_modl_d}
\end{align}
\end{subequations}
where the rate threshold $\bar{r}$ represents the QoS constraints.

Note that the problem in (\ref{ee_modl}) is nonconvex because of the nonconcave objective functions (\ref{ee_modl_a}) and the nonlinear constraint (\ref{ee_modl_c}).
In the next section, we propose a novel optimization scenario for solving problem (\ref{ee_modl}) in real-time embedded application.

\section{Joint harvesting time and power allocation (JHTPA) for EE maximization}
In this section, we propose a practical algorithm for the EE maximization problem (\ref{ee_modl}) by jointly optimizing  the energy harvesting time and power allocation. To solve the problem (\ref{ee_modl}), we first change the variables \cite{Nasir_TCOM_2017}
\[
1 - \tau = \frac{1}{\theta} \ \mbox{and} \ p_n \rightarrow \frac{1}{p_n}, \ n = 1,..., N
\]
such that the variable satisfy the convex constraint
\begin{align} \label{opt_time_var}
\theta > 1.
\end{align}

Then, the problem (\ref{ee_modl}) is equivalent to
\begin{subequations} \label{ee_modl_2}
	\begin{align}
	{\max\limits_{\theta,\bp}}& \ \phi = \frac{\sum_{n=1}^N r_n(\theta, \bp)}{\vartheta(\theta, \bp)} \label{ee_modl_2a} \\
	\mbox{s.t.} & \ (\ref{opt_time_var}), \\
	& \ {1}/{ p_n} \leq (\theta - 1) \eta P_0 g_n, \ n \in \mathcal{N} \label{ee_modl_2b} \\
	& \ \frac{1}{\theta} \ln \left(1 + \frac{h_{n,n}}{p_n \sum_{i \neq n}^{N} h_{n,i}/p_i + p_n  \sigma^2}\right) \geq \bar{r} , \ n \in \mathcal{N}. \label{ee_modl_2c}
	\end{align}
\end{subequations}
where $\vartheta(\theta, \bp) = \sum_{n=1}^{N} 1/(\theta p_n) + (1 - 1/\theta) \eta P_0 + P_{\sf cir}$.

To solve the problem (\ref{ee_modl_2}), we use the logarithmic inequality \cite{nguyenTSP2017}
\begin{eqnarray} \label{ineq_1}
\frac{1}{t}\ln(1+\frac{1}{xy}) \geq \frac{2}{\bar{t}} \ln\Big(1+\frac{1}{\bar{x} \bar{y}}\Big) + \frac{2}{\bar{t}(\bar{x}\bar{y}+1)} 
-\frac{1}{\bar{t}\bar{x}(\bar{x}\bar{y}+1)} x -\frac{1}{\bar{t}\bar{y}(\bar{x}\bar{y}+1)}y
 -\frac{\ln(1+{1}/{\bar{x}\bar{y}})}{\bar{t}^2} t \nonumber \\
 \forall t>0, \bar{t}>0, x>0, \bar{x}>0, y>0, \bar{y}>0 .
\end{eqnarray}
which follows from the convexity of function $\ln \Big(1+{1}/{xy}\Big)/t$.

For
\[ 
x = p_n/h_{n,n}, y=\sum_{i \neq n}^{N} h_{i,n}/p_i + \sigma^2, t=\theta,
\]
\[
\bar{x} = x^{(\kappa)} = p_n^{(\kappa)}/h_{n,n}, \bar{y} = y^{(\kappa)} =\sum_{i \neq n}^{N} h_{i,n}/p_i^{(\kappa)} + \sigma^2, \bar{t}=t^{(\kappa)}=\theta^{(\kappa)},
\]
thus, the throughput can be approximated as
\begin{align}
r_n(\theta, \bp) \geq \psi_n^{(\kappa)}(\theta,\bp)
\end{align}
where
\begin{eqnarray} \label{ee_modl_ineq}
\psi_n^{(\kappa)}(\theta,\bp) =
\frac{2}{t^{(\kappa)}} \ln\Big(1+\frac{1}{x^{(\kappa)} y^{(\kappa)}}\Big) + \frac{2}{t^{(\kappa)}(x^{(\kappa)}y^{(\kappa)}+1)} 
-\frac{1}{t^{(\kappa)}x^{(\kappa)}(x^{(\kappa)}y^{(\kappa)}+1)} x \nonumber \\
 -\frac{1}{t^{(\kappa)}y^{(\kappa)}(x^{(\kappa)}y^{(\kappa)}+1)} y -\frac{\ln(1+{1}/{x^{(\kappa)}y^{(\kappa)}})}{(t^{(\kappa)})^2} t.
\end{eqnarray}

With the feasible points ($\theta^{(k)}$, $\bp^{(k)}$) of (\ref{ee_modl_2}), one has
\[
\phi^{(\kappa)} = \sum_{n=1}^{N} \psi_n(\theta^{(\kappa)},\bp^{(\kappa)})/\vartheta(\theta^{(\kappa)}, \bp^{(\kappa)}).
\]

At the $\kappa$th iteration, the following convex program is solved to generate the next feasible point
\begin{subequations} \label{ee_modl_3}
	\begin{align}
	{\max\limits_{\theta,\bp}}&\ \ \sum_{n=1}^N \psi_n^{(\kappa)}(\theta,\bp) - \phi^{(\kappa)} \vartheta^{(\kappa)}(\theta, \bp)  \label{ee_modl_3a} \\
	\mbox{s.t.} & \ (\ref{opt_time_var}), (\ref{ee_modl_2b}),  \\
	& \ \psi_n^{(\kappa)}(\theta,\bp) \geq \bar{r} , \ n \in \mathcal{N}. \label{ee_modl_3c}
	\end{align}
\end{subequations}
where $\vartheta^{(\kappa)}(\theta, \bp) = \sum_{n=1}^{N} 1/(\theta p_n) + (1 - 2/\theta^{(\kappa)} + \theta/(\theta^{(\kappa)})^2) \eta P_0 + P_{\sf cir}$.

We propose an algorithm to solve the EE maximization (\ref{ee_modl_3}). The initial point ($\theta^{(0)}$, $\bp^{(0)}$) for (\ref{ee_modl_3}) is easily located by random search such that it satisfies the constraints in problem (\ref{ee_modl_2}).

\begin{algorithm}[H]
	\caption{: Joint optimal harvesting time and power allocation problem (\ref{ee_modl_2})} \label{alg1}
	\begin{algorithmic}[1]
		\State \textbf{Initialization}: Set feasible points $\theta^{(0)}$, $\bp^{(0)}$, $\kappa=0$ and $\phi^{(0)} = \sum_{n=1}^{N} \psi_n(\theta^{(0)},\bp^{(0)})/\vartheta(\theta^{(0)}, \bp^{(0)})$. Set the tolerance $\varepsilon=10^{-2}$. 
		\State \textbf{Repeat}
		\State \quad Solve the (\ref{ee_modl_3}) for the optimal solution $(\theta^{(\kappa+1)},\bp^{(\kappa+1)})$. Set $\phi^{(\kappa+1)} = \sum_{n=1}^{N} \psi_n(\theta^{(\kappa+1)},\bp^{(\kappa+1)})/\vartheta(\theta^{(\kappa+1)}, \bp^{(\kappa+1)})$.
		\State Set $\kappa:=\kappa+1$
		\State \textbf{Stop} convergence of the objective in (\ref{ee_modl_3}). 
	\end{algorithmic}
\end{algorithm}

\section{Near-optimal resource allocation algorithms for EE maximization}
In this section, two low-complexity procedures are presented as conventional methods to evaluate the effectiveness of JHTPA in EE performance and solving time.
\subsection{Optimal Power allocation (OPA)}
This algorithm addresses power allocation for EE maximization problem (\ref{ee_modl}) where the harvesting time value is fixed as $ 1 - \tau = 1/{\theta_{\sf fix}}$, ${\theta_{\sf fix}} > 1$. Thus, problem (\ref{ee_modl}) is equivalent to
\vspace{-0.5cm}
\begin{subequations} \label{ee_modl_PA}
	\begin{align}
	{\max\limits_{\bp}}& \ \phi = \frac{\sum_{n=1}^N r_n(\theta_{\sf fix}, \bp)}{\vartheta(\theta_{\sf fix}, \bp)} \label{ee_modl_PAa} \\
	\mbox{s.t.} & \ {p_n} \leq ({\theta_{\sf fix}} - 1) \eta P_0 g_n, \ n \in \mathcal{N} \label{ee_modl_PAb} \\
	& \ \ln \left(1 + \frac{p_n h_{n,n}}{\sum_{i \neq n}^{N} h_{n,i} p_i + \sigma^2}\right) \geq  {\theta_{\sf fix}} \bar{r}, \ n \in \mathcal{N}. \label{ee_modl_PAc}
	\end{align}
\end{subequations}
where $\vartheta(\theta_{\sf fix}, \bp) = \sum_{n=1}^{N} p_n/\theta_{\sf fix} + (1 - 1/\theta_{\sf fix}) \eta P_0 + P_{\sf cir}$.

To solve the nonconvex problem (\ref{ee_modl_PA}), we apply the inequality (\ref{ineq_1}) for
\[
x = 1/p_n h_{n,n}, \ y=\sum_{i \neq n}^{N} h_{i,n} p_i + \sigma^2, \ t = 1, \
\]
and
\[
\bar{x} = x^{(\kappa)} = 1/p_n^{(\kappa)} h_{n,n}, \
\bar{y} = y^{(\kappa)} =\sum_{i \neq n}^{N} h_{i,n}p_i^{(\kappa)} + \sigma^2, \
\bar{t} = t^{(\kappa)} = 1.
\]

Then, the numerator of objective function in (\ref{ee_modl_PA}) can be approximated as
\begin{align}
r_n(\theta_{\sf fix}, \bp) \geq \bar{\psi}_n^{(\kappa)}(\theta_{\sf fix},\bp)
\end{align}
where $\bar{\psi}_n^{(\kappa)}(\theta_{\sf fix},\bp)$ is defined as (\ref{ee_modl_ineq}).

At the $\kappa$th iteration, the following convex program is solved to generate the next feasible point
\vspace{-0.7cm}
\begin{subequations} \label{ee_modl_PA_1}
	\begin{align}
	{\max\limits_{\bp}} &\ \ \sum_{n=1}^N \bar{\psi}_n^{(\kappa)}(\theta_{\sf fix},\bp) - \phi^{(\kappa)} \vartheta(\theta_{\sf fix}, \bp)  \label{ee_modl_PA_1a} \\
	\mbox{s.t.} & \ (\ref{ee_modl_PAb}), (\ref{ee_modl_PAc}) \label{ee_modl_PA_1b}
	\end{align}
\end{subequations}
where $\phi^{(\kappa)} = \sum_{n=1}^{N} \bar{\psi}_n(\theta_{\sf fix}, \bp^{(\kappa)}) / \vartheta(\theta_{\sf fix}, \bp^{(\kappa)})$.

\subsection{Optimal harvesting time (OHT)}
This algorithm solves the harvesting time optimization problem in the slack variable of $\theta$ with the use of maximum harvested power in D2D communication as follows:
\[
p_{n} = (\theta-1) \eta P_0 g_n
\]

Therefore, the maximin sum-rate problem with fixed harvested energy is given by
\begin{eqnarray} \label{ee_modl_TH}
{\max\limits_{\theta}} \ {\min\limits_{n \in \mathcal{N}}} \ {r_n(\theta)} \label{ee_modl_THa} \quad \mbox{s.t.}  \ (\ref{opt_time_var}) \label{ee_modl_THb}
\end{eqnarray}
where $r_n(\theta) = \frac{1}{\theta} \ln \left(1 + \frac{(\theta-1) h_{n,n}  g_n }{(\theta-1) \sum_{i \neq n}^{N} h_{n,i} g_i + \sigma^2/ \eta P_0}\right)$.

Next, the objective function in (\ref{ee_modl_TH}) can be approximated by using the inequality (\ref{ineq_1}) for
\[
x = 1 / (\theta-1) h_{n,n} g_n, \ y = (\theta-1) \sum_{i \neq n}^{N} h_{i,n} g_i + \sigma^2/\eta P_0, \ t = \theta, \
\]
and
\[
\bar{x} = x^{(\kappa)} = 1/ (\theta^{(\kappa)}-1) h_{n,n} g_n, \
\bar{y} = y^{(\kappa)} = (\theta^{(\kappa)}-1) \sum_{i \neq n}^{N} h_{i,n}g_i + \sigma^2/ \eta P_0, \ \bar{t} = t^{(\kappa)} = \theta^{(\kappa)},
\]
and thus one has
\begin{align}
r_n(\theta) \geq \hat{\psi}_n^{(\kappa)}(\theta)
\end{align}
where $\hat{\psi}_n^{(\kappa)}(\theta)$ is defined as (\ref{ee_modl_ineq}).

At the $\kappa$th iteration, the following max-min program is solved to generate the next feasible point
\begin{eqnarray} \label{ee_modl_TH_1}
	{\max\limits_{\theta}} \ {\min\limits_{n \in \mathcal{N}}} \ \hat{\psi}_n^{(\kappa)}(\theta) \quad \mbox{s.t.}  \ (\ref{opt_time_var}).
\end{eqnarray}

Hence, we assume that $\theta^*$ is the optimal solution of problem (\ref{ee_modl_TH}). Then, the EE performance is defined as
\begin{equation}
\phi(\theta^*) = \frac{\sum_{n=1}^N r_n(\theta^*)}{\vartheta(\theta^*)}
\end{equation}
where $\vartheta(\theta^*) = (1 - 1/\theta^*)\eta P_0 (\sum_{n=1}^{N} g_n + 1) + P_{\sf cir}$.

\section{Implementations}
In this section, we evaluate the performance of the UAV network by embedded optimization module implemented in Python \cite{cvxpy}. The results are obtained using CVXPY 0.4.11 package with ECOS solver. The computational platform is a laptop with an Intel Core(TM) i7, CPU @2.80GHz and 16GB memory.

\begin{figure}[H]
	\centering
	\centerline{\includegraphics[trim=0cm 0cm 0cm 0.8cm, width=0.6\textwidth]{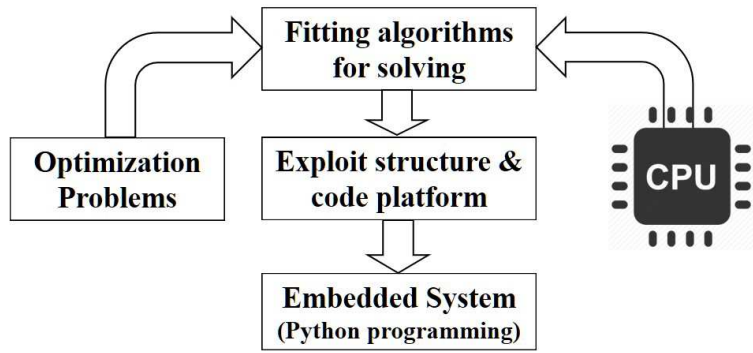}}
	\caption{A structure of real-time embedded optimization.}
	\label{fig:Diagram}
\end{figure}

An example structure of real-time embedded optimization for UAV network is shown in Fig. \ref{fig:Diagram}. We consider a center unit (CU) as a ground station for exchanging information of D2D netowrk with UAV and the UAV mobile is located at the center of a circle coverage network with radius $800$m while being at the height of $H=50$. The D2D pairs are randomly distributed in the coverage network and the maximum distance between D2D-Tx and D2D-Rx is $50$m.

Similar to \cite{Wang2017TGCN}, the channel power gain between D2D-Tx and D2D-Rx is modelled as
\begin{eqnarray} \label{chan_D2D}
h_{n,n} &=& \beta_0 \rho_n^2 D^{-\alpha_h}
\end{eqnarray}
where $\beta_0$ is the channel power gain at the reference distance $d_0$, $\rho_n$ is an exponentially distributed random variable with unit mean, $D$ is the distance between D2D-Tx and D2D-Rx, and $\alpha_h$ represents the path loss exponent for D2D links.

Furthermore, we exploit the air-to-ground (ATG) channel model for D2D UAV-assisted communication \cite{Holis2008TAP, Mozaffari2016TWCOM}. The channel power gain from the UAV to the $n$th D2D-Tx located at $(x, y)$ under the LOS or NLOS links is given by
\begin{eqnarray} \label{chan_UAV_D2D}
g_n &=& Pr_{LOS} \times (\sqrt{x^2 + y^2 + H^2})^{-\alpha_g} + Pr_{NLOS} \times \gamma (\sqrt{x^2 + y^2 + H^2})^{-\alpha_g} \quad
\end{eqnarray}
where $Pr_{LOS} = 1/(1+a \times {\sf exp}(-b[\varphi - a]))$ is the LOS probability where $a$ and $b$ are constant values depending on the environment. Then, one has $Pr_{NLOS} = 1 - Pr_{LOS}$, and $\alpha_g$ represents the path loss exponent from UAV to D2D-Tx. The elevation angle $\varphi$ in terms of degree unit is given by $\varphi = \frac{180}{\pi} \times {\sf sin}^{-1} \left( \frac{H}{\sqrt{x^2+y^2+H^2}} \right)$.

The QoS constraint is set as
\begin{eqnarray} \label{QoS_thres}
\bar{r} &=& {\sf min} \{ r_n(\theta_{\sf fix}), 0.2 \} {\sf bps/Hz}
\end{eqnarray}
where $r_n(\theta_{\sf fix})$ is defined in (\ref{ee_modl_TH}).

Other simulation parameters in the considered D2D UAV network are provided in Table \ref{Chan_model} as \cite{Wang2017TGCN, Mozaffari2016TWCOM}.
\begin{table}[ht]
	\centering
	\caption{Simulation parameters \label{Chan_model}}
	\begin{tabular}{ |p{5.5cm}||p{4cm}| }
		\hline
		Parameter & Numerical value \\
		\hline
		Bandwidth   							& $1$ MHz  \\	
		UAV transmission power					& $5$ W \\
		Path-loss exponents						& $\alpha_h = 3$, $\alpha_g = 3$ \\
		Channel power gain at the reference  	& $\beta_0 = -30$ dB \\
		Noise power density 					& $-130$ dBm/Hz \\
		Energy harvesting efficiency 			& $\eta = 0.5$ \\
		UAV non-transmission power 				& $4$ W \\
		ATG channel parameters					& $a = 11.95$, $b = 0.136$ \\
		The excessive attenuation factor		& $\gamma = 20$ dB \\
		\hline
	\end{tabular}
\end{table}
\begin{figure}[H]
	\centering
 	\centerline{\includegraphics[trim=0cm 0.5cm 0cm 1.5cm, width=0.6\textwidth]{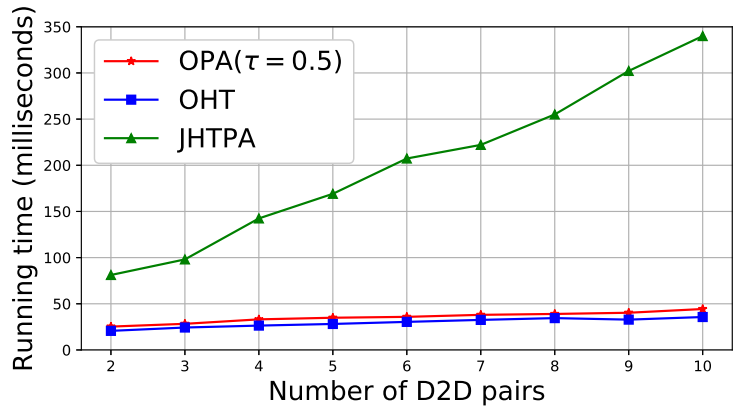}}
	\caption{The average running times of OPA for $\tau=0.5$, OHT and JHTPA algorithms versus the number of D2D pairs.}
	\label{fig:time_solve}
\end{figure}
Fig. \ref{fig:time_solve} plots the average running time for solving the algorithms of JHTPA, OPA, OHT. As can be observed from this figure, the solving time of all algorithms is in milliseconds within $10$ D2D pairs. For instance, with $5$ D2D pairs, the running time is lower than $50$ milliseconds for OPA and OHT and around $150$ milliseconds for JHTPA.

\begin{figure}[H]
	\centering
	\centerline{\includegraphics[trim=0cm 0.5cm 0cm 1cm, width=0.6\textwidth]{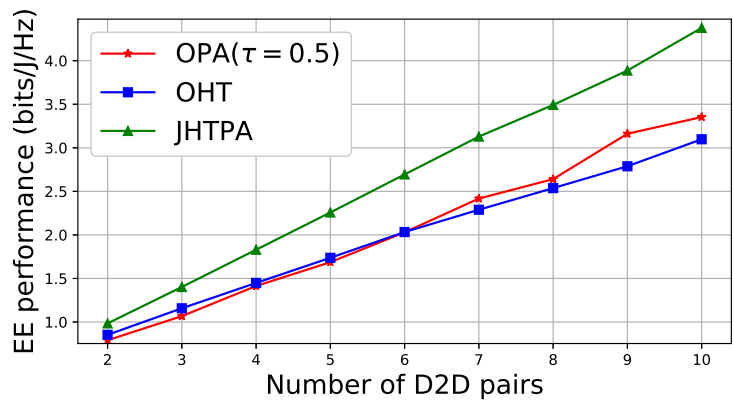}}
	\caption{The EE performance of OPA for $\tau=0.5$, OHT and JHTPA algorithms versus the number of D2D pairs.}
	\label{fig:ee_solve}
\end{figure}

From Figs. \ref{fig:time_solve} and \ref{fig:ee_solve}, we demonstrate the trade-off between the solving time and the EE performance which should be carefully considered in real time applications. Although the running time for JHTPA algorithm is higher than that for OPA and OHT, the EE performance in JHTPA significantly outperforms the two other algorithms. Interestingly, the EE performance of OPA algorithm exceeds that for OHT algorithm when the number of D2D pairs increases, while the running time in these two algorithms is almost identical. As such, by focusing on adaptive power allocation, the OPA algorithm offers a better solution over the OHT algorithm.

\section{Conclusions}
In this paper, we have proposed the real-time resource allocation for D2D communications assisted by UAV. We have shown that our real-time optimization is very suitable for UAV application where the real-time control is a crucial issue.



\bibliographystyle{IEEEtran}
\bibliography{WCL2018-0585}

\end{document}